\documentclass[preprint,pteplogo]{ptephy} 
\preprintnumber{KUNS-2568  / YITP-15-58} 

\usepackage{color}
\usepackage{amsmath}
\usepackage{graphicx}
\usepackage{amssymb}
\usepackage{bm}

\newcommand{\Comments}[1]{}

\newcommand{\be}{\begin{equation}}
\newcommand{\ee}{\end{equation}}
\newcommand{\ba}{\begin{align}}
\newcommand{\ea}{\end{align}}
\newcommand{\nn}{\nonumber}

\newcommand{\chibar}{{\bar{\chi}}}

\newcommand{\Nc}{N_c}
\newcommand{\non}{\nonumber}
\newcommand{\hatmu}{\hat{\mu}}
\newcommand{\Fint}{{\cal D}}
\newcommand{\Seff}{S_\mathrm{eff}}

\newcommand{\bk}{\mathbf{k}}
\newcommand{\CUM}[1]{\chi_\mu^{(#1)}}
\newcommand{\RX}{R(\bold{x})}
\usepackage[normalem]{ulem}  
\newcommand{\com}[1]{{\color[rgb]{0,0,1}{#1}}}

\renewcommand\sout{\bgroup \color{red} \ULdepth=-.5ex \ULset}
\newcommand\soutb{\bgroup \color{blue} \ULdepth=-.5ex \ULset}

\newcommand{\vk}{{\bold{k}}}

\newcommand{\vkt}{{\vk,\tau}}
\newcommand{\vkmt}{{-\vk,\tau}}

\newcommand{\expv}[1]{\left< #1 \right>}
\newcommand{\mrm}[1]{\mathrm{#1}}
\newcommand{\Nt}{N_{\tau}}
\newcommand{\equref}[1]{Eq.~(\ref{#1})}
\newcommand{\secref}[1]{Sec.~\ref{#1}}
\newcommand{\figref}[1]{Fig.~\ref{#1}}

\newcommand\soutg{\bgroup \color{green} \ULdepth=-.5ex \ULset}

\renewcommand{\com}[1]{{\color[rgb]{0,0,0}{#1}}}
\renewcommand\soutg{\bgroup \color{red} \ULdepth=-.5ex \ULset}

\begin{document}

\title{Net baryon number fluctuations across the chiral phase transition
at finite density in the strong coupling lattice QCD}

\author{
\name{Terukazu Ichihara}{1,2,\ast}, 
\name{Kenji Morita}{2,3,\ddag}
and \name{Akira Ohnishi}{2,\dag}
}


\address{\affil{1}{Department of Physics, Kyoto University,
         Kyoto 606-8502, Japan}
\affil{2}{Yukawa Institute for Theoretical Physics, Kyoto University,
         Kyoto 606-8502, Japan}
\affil{3}{Institute of Theoretical Physics, University of Wroclaw,
         PL-50204 Wroc\l aw, Poland}
\email{t-ichi@ruby.scphys.kyoto-u.ac.jp}, 
{\dag}ohnishi@yukawa.kyoto-u.ac.jp, 
{\ddag}kmorita@yukawa.kyoto-u.ac.jp
}


\begin{abstract}%
We investigate the net-baryon number fluctuations
across the chiral phase transition at finite density
in the strong coupling and chiral limit.
Mesonic field fluctuations are taken into account
by using the auxiliary field Monte-Carlo method.
We find that the higher-order cumulant ratios,
$S\sigma$ and $\kappa\sigma^2$, show oscillatory behavior 
around the phase boundary at $\mu/T\gtrsim 0.2$,
and there exists the region
where the higher-order cumulant ratios are negative.
The negative region of $\kappa\sigma^2$ is found to shrink with increasing lattice size.
This behavior agrees with the expectations from the scaling analysis.
\end{abstract}

\subjectindex{ B64, 
      D28, 
      D30, 
      D34 
      }

\maketitle
\section{Introduction}
\label{sec:Intro}

Fluctuations of conserved charges are promising observables in
relativistic heavy-ion collisions in search for QCD phase transition
\cite{SRS1999,AHM-JK2000,Koch_review}.
In the first phase of the Beam Energy Scan (BES I) program at Relativistic
Heavy Ion Collider (RHIC), net-proton \cite{BESI-proton},
as a proxy of net-baryon \cite{Hatta-Stephanov,KA_proton-baryon},
and net-electric charge event-by-event fluctuations \cite{BES1-charge}
have been measured in Au+Au collisions for a broad energy range from
$\sqrt{s_{NN}}=7.7$ to 200 GeV. 
The success of the statistical thermal model \cite{Thermalmodel}
indicates the event-by-event
fluctuations of the multiplicity of conserved charges can be regarded as
particle number fluctuations in the grand canonical ensemble specified
by temperature, volume and chemical potential. \footnote{This argument
holds for a limited acceptance. If one covers entire phase space in heavy
ion collisions, those fluctuations is of course absent. One can invoke
more informations by changing the rapidity acceptance beyond the equilibrium
regime \cite{Kitazawa_PLB728}.
}

The particle number fluctuations of the conserved charge, i.e.,
susceptibilities or cumulants, reflect the property of the phase in QCD
\cite{SFR_susceptibility},
which is expected to exhibit a rich structure \cite{Fukushima-Hatsuda}.
Especially, if QCD has a critical point (CP) \cite{QCDCP},
the susceptibilities and higher-order cumulants
diverge at the critical point owing to the divergent correlation 
length~\cite{Hatta-Ikeda,Stephanov_higherorder}, 
thus one expects anomalously large
fluctuations would be observed if the
system passes around CP. 
In heavy-ion collisions, the correlation length cannot diverge due to
finite size of the system, but these fluctuations are expected to remain
sensitive to the remnant 
criticality of the system 
around 
CP~\cite{SRS1999}.
Since this sensitivity implies the tail of
the event-by-event multiplicity distribution has importance in the
behavior of the cumulants~\cite{Morita_prob},
the analysis becomes statistically demanded.

The measured higher-order cumulant ratios of net-proton number,
$S\sigma=\chi_p^{(3)}/\chi_p^{(2)}$
and $\kappa\sigma^2=\chi_p^{(4)}/\chi_p^{(2)}$,
show non-monotonic behavior
as a function of the incident energy,
where $\sigma^2$, $S$ and $\kappa$ are referred to 
as the variance, {\em skewness} and {\em kurtosis},
respectively.
At around $\sqrt{s_{\scriptscriptstyle NN}}=20~\mathrm{GeV}$,
the data show $\kappa\sigma^2 < 1$ below the expected value 
in the Skellam distribution \cite{BESI-proton}.

The decrease of $\kappa\sigma^2$ in the net-proton
number can be the signal of the critical behavior.
According to the theoretical arguments \cite{SF-4-CM,Stephanov_negativekappa}, 
$\kappa\sigma^2$ of the net-baryon number can be reduced by critical behavior from universality. 
In QCD, the expected universality class depends on the quark masses.
For two-flavor massless quarks with axial anomaly,
the QCD phase transition 
at finite $T$ ($\mu=0$)
belongs to the universality class of 
3d O(4) symmetric spin model~\cite{PW84,MagneticEOS},
and the second order transition at low $\mu$ may turn into the first order
transition at the tricritical point (TCP).
At physical quark masses,
finite $T$ phase transition would be crossover~\cite{FSS},
and the fluctuations are
governed by the approximate chiral symmetry.
The pseudo-critical line at low $\mu$ may be connected 
with the first order transition line at CP,
whose criticality is expected to belong to Z(2) universality 
class~\cite{QCDCP,Hatta-Ikeda,Fujii-Ohtani}.

While the universality argument dictates the singular behavior of the
thermodynamic quantities close to
the phase transition,
the actual magnitude of the fluctuations are smeared
by the finite volume effect 
in addition to the finite quark mass.
Finite volume makes the transition smoother.
We need to perform finite size scaling analysis
to observe precise critical behavior \cite{FSS}.
The critical behavior depends on
the relative strength of the singular part to the regular (non-singular) part  
of the free energy density (or its derivatives).
The regular part would also mask
the expected singular behavior from criticality \cite{SF-4-CM}.
Thus, explicit calculations are desirable
to see how much criticality can be present in the fluctuation 
observables,
and to pin down the origin of 
the observed decrease of $\kappa\sigma^2$:
Z(2) critical behavior around CP \cite{Stephanov_higherorder}, 
the remnant of the O(4) chiral phase transition~\cite{SF-4-CM},
or other mechanisms.

Lattice QCD (LQCD) is the most powerful non-perturbative approach to QCD
based on Monte-Carlo (MC) simulations,
and is successful at vanishing or low baryon chemical potential.
For instance,
higher-order cumulants have been studied at zero chemical potential \cite{HM-Lattice-zero-chemi-pote,Jin:2014hea,Allton:2005gk} and at finite chemical potential \cite{HM-lattice,Jin:2013wta,Allton:2005gk,Takeda:2014nta,Gattringer:2014hra}.
At large baryon chemical potential, however,
LQCD faces the notorious sign problem 
which makes it difficult to carry out MC simulations
owing to complex fermion determinant.
There are many attempts to circumvent the sign 
problem~\cite{Reweighting,Taylor,Histogram,ImagMu,Langevin,LThimble,Fugacity,Canonical}.
Most of the methods evading the sign problem are reliable
only in limited circumstances such as $\mu /T \lesssim 1$,
small volume, or heavier quark masses than physical one,
depending on the method.
Conclusive results at physical point have not been obtained yet.

Owing to the sign problem in 
 LQCD,
most of fluctuation studies 
at nonzero density have
been carried out by using chiral models~\cite{CModel-3rd,CModel-HM,FRG-HM,SF-4-CM,Morita_prob, Stokic, Morita_ptep}.
It has been pointed out that taking
into account fluctuations around the phase transition is essential to
correctly describe the behavior of the fluctuation of conserved
charges \cite{FRG-HM,SF-4-CM,Morita_prob}. 
In chiral models, this has been done by implementing functional
renormalization group (FRG) \cite{FRG, Schaefer-Wambach}. 
This is a natural consequence of the fact that the critical behavior of
the conserved charge fluctuations is governed by the critical exponent
of the specific heat $\alpha$ which requires beyond mean-field treatment
\cite{FRG-HM,SF-4-CM,Morita_prob, Stokic, Morita_ptep}.

One of the possible alternative ways to attack the finite density region is 
the strong coupling approach
of LQCD with fermions
\cite{SCL,large_d,Faldt,BilicDeme,MF-SCL,Bilic,Jolicoeur,NLOchiral,Aoki:1987us,NLOPD,NNLO,KarschMutter,MDP,MDP-CE,SC-Rewei,SC-Pol,Ichihara:2014ova,PhDFromm}.
The strong coupling approach with fermions could reduce the severity of the sign problem
and is applied
to investigate the chiral phase transition.
Recently, theoretical frameworks including fluctuation effects in the strong coupling limit (SCL) have been developed:
auxiliary field Monte-Carlo (AFMC) method
 \cite{Ichihara:2014ova}
and the
monomer-dimer-polymer (MDP) simulation~\cite{MDP,MDP-CE,SC-Rewei, KarschMutter,PhDFromm}. 
These two methods give consistent
phase boundaries of the chiral phase transition
 in the chiral limit.
Although SCL is the coarse lattice limit,
we might expect that the observables
related to the phase transition is not so sensitive to the coarse
lattice spacing since long-wave dynamics dominates the property of the phase
transition, including the influences on the higher-order 
cumulants~\cite{Morita_ptep}.
Furthermore, 
one can take the chiral limit in which the chiral transition becomes
second order. Thus, the singular behavior associated with the phase
transition is smeared by the finite size effect only.
As a result, the divergent part
of the higher-order cumulants, which can
be described by relevant scaling function of the singular part of the
free energy density \cite{SF-4-CM}, 
could be replaced by sign changes across
the transition \cite{CModel-3rd,FRG-HM,SF-4-CM,Stephanov_negativekappa}. 

In this article,
we focus on the fluctuations of net-baryon number and
discuss the higher-order cumulant ratios  
in the strong coupling limit of 
LQCD 
by utilizing the AFMC method to take the fluctuation effects into account.
We here consider the LQCD action with one species of unrooted staggered fermion
in the chiral limit.
We demonstrate that the higher-order cumulant ratios,
$S\sigma$ and $\kappa\sigma^2$, show oscillatory behavior,
and there exists the region where higher-order cumulants are negative
in the strong coupling and chiral limit on not too small lattices
at $\mu/T\gtrsim 0.2$.
We also discuss the lattice size dependence of the negative region.

This paper is organized as follows.
We briefly
provide the formalism in
\secref{sec:EA}. 
Our main results on the higher-order fluctuations of the net-baryon
number will be presented in \secref{sec:results}.
We summarize our paper in \secref{sec:summary}.
Some formulae for the net-baryon number cumulants 
used in \secref{sec:results} are found
in the appendix.

\section{Lattice QCD in the strong coupling limit with fluctuations}
\label{sec:EA}
We consider a lattice QCD action with one species of unrooted
staggered fermion 
in the $d (= 3) + 1$
dimensional anisotropic Euclidean spacetime with $\Nt$ temporal and $L$
spatial lattice sizes.
This LQCD action has remnant chiral symmetry $U(1)_R \times U(1)_L$
in the chiral limit ($m_0 \rightarrow 0$).
We set the number of the color $N_c = 3$ and the lattice unit $a=1$ throughout this paper. 
In the following, we briefly summarize the formalism developed in Ref.~\cite{Ichihara:2014ova}. 

\subsection{Effective action in the strong coupling limit with fluctuations}
\label{subsec:Latac}
In the strong coupling limit (SCL) $g\rightarrow \infty$, we could ignore the plaquette terms,
which are proportional to $1/g^2$, so the partition function and the
action of the lattice QCD become
\begin{align}
  {\cal Z}_{\mathrm{SCL}} 
  =&  \int \Fint \left[ \chi,\chibar,U_\nu \right] e^{-S_\mathrm{F}}
  \ ,\\
  S_F
  =&\frac12 \sum_x \left[
  	V^{+}_x - V^{-}_x
  		\right]
  +\frac{1}{2} \sum_{x}\sum_{j=1}^{d} \eta_{j,x}\left[
  		 \bar{\chi}_x U_{j,x} \chi_{x+\hat{j}}
  		-\bar{\chi}_{x+\hat{j}} U^\dagger_{j,x} \chi_{x}
  		\right]
  +m_0 \sum_{x} M_x
\label{Eq:LQCD}
\ ,\\
V^{+}_x=&\gamma e^{\mu/f(\gamma)} \bar{\chi}_x U_{0,x} \chi_{x+\hat{0}}
\ ,\quad
V^{-}_x=\gamma e^{-\mu/f(\gamma)} \bar{\chi}_{x+\hat{0}} U^\dagger_{0,x} \chi_x
\ ,\quad
M_x=\bar{\chi}_x \chi_x
\ ,
\end{align}
where $\chi_x$ and $U_{\nu,x}$
represent the staggered quark field and the link variable, 
respectively,
and $V^{\pm}_x$ and $M_x$ are mesonic composites.
The staggered sign factor
$\eta_{j,x}=(-1)^{x_0+\cdots+x_{j-1}}$
is related to the gamma matrices in the continuum limit.
The quark mass $m_0$ is taken to be zero throughout this paper.
Quark chemical 
potential $\mu$ is introduced together with
the physical lattice spacing ratio 
$f(\gamma)=a_\mrm{s}^\mrm{phys}/a_\tau^\mrm{phys}$, %
where $\gamma$ is %
the anisotropy parameter.
We adopt $f(\gamma)=\gamma^2$
following the arguments in Refs.~\cite{Bilic,BilicDeme,PhDFromm,Ichihara:2014ova}

We obtain the effective action of the auxiliary fields, $S_\mathrm{eff}^\mathrm{AF}$, after
the following three steps.
First, by integrating out spatial link variables
\cite{SCL,large_d,MF-SCL,Faldt,BilicDeme,Bilic,NLOPD,NNLO,Jolicoeur}
in the leading order of the strong coupling and
$1/d$ (large-dimensional) expansion~\cite{large_d},
one finds a convenient expression for the effective action,
\begin{align}
S_\mathrm{eff}
&=\frac12 \sum_x \left[ V^{+}_x - V^{-}_x \right]
- \displaystyle \frac {1}{4N_c} \sum_{x,j} M_x M_{x+\hat{j}}
+m_0 \sum_{x} M_x
\label{Eq:Seff}
\ .
\end{align}
It should be noted that spatial baryonic hopping terms are not included
in \equref{Eq:Seff}
as they are higher order in the $1/d$ expansion. 
By comparison,
we exactly integrate out temporal link variables later, then
the temporal baryonic hopping effects are taken into account.%

In the second step, we transform the effective action
to the fermion-bilinear form by using
the extended Hubbard-Stratonovich (eHS) transformation
\cite{NLOPD,NNLO}, 
\begin{align}
e^{\alpha A B}  
&= \int\, d\varphi\, d\phi\,
	e^{-\alpha\left\{
        \left[
		 \varphi-(A+B)/2
        \right]^2
        +\left[
		 \phi - i(A-B)/2
        \right]^2
		\right\}+ \alpha AB}
\nn \\
&= \int\, d\psi\, d\psi^*\,
	e^{-\alpha\left\{
	\psi^* \psi-{A}\psi-\psi^* B
	\right\}}
\ ,\label{Eq:EHSp}
\end{align}
where $\psi=\varphi +i \phi$ and $d\psi\,d\psi^*=d\mathrm{Re}\psi\,d\mathrm{Im}\psi=d\varphi d\phi$.
The four-Fermi interaction terms 
in \equref{Eq:Seff} are diagonal in the momentum space, and are separated
into two parts based on the momentum regions: 
the positive modes ($f(\bold{k})=\sum_{j=1}^{d} \cos\, k_j > 0$) and the negative ($f(\bold{k}) < 0$) modes.
\begin{align}
- \displaystyle \frac {1}{4N_c} \sum_{x,j} M_x M_{x+\hat{j}} 
&= -\frac{L^3}{4N_c}
	\sum_{\bold{k}, \tau}
	f(\bold{k})\, M_{-\bold{k},\tau}\, M_{\bold{k},\tau}
\ \nonumber\\
&=- \frac {L^3}{4N_c} \sum_{\bk, \tau, f(\bold{k})>0} f(\bold{k}) 
(M_{\bk,\tau} M_{-\bk,\tau}
-M_{\bar{\bk},\tau} M_{-\bar{\bk},\tau})
\ ,
\label{Eq:SsFT}
\end{align}
where 
$M_{x=(\bold{x},\tau)}=\sum_{\bold{k}} e^{i\bold{k}\cdot\bold{x}} M_{\bold{k},\tau}$,
$\bar{\bk}=\bk+(\pi,\pi,\pi)$,
and $f(\bar{\bold{k}})=-f(\bold{k})$.
The effective action after the eHS transformation of Eq.~\eqref{Eq:SsFT} reads
\begin{align}
S_\mathrm{eff}^\mathrm{EHS}
=&\frac{1}{2}\sum_x\left[V_x^+ - V_x^-\right]
 +\sum_x m_x M_x
+\frac{L^3}{4N_c} \sum_{\vkt, f(\vk)>0}
f(\vk)\left[\left|\sigma_\vkt\right|^2+\left|\pi_\vkt\right|^2\right]
\label{Eq:SeffEHS}
\ ,
\\
m_x
=&
 m_0
 +\frac{1}{4N_c} \sum_{j}
	\left[
	 (\sigma+i\varepsilon\pi)_{x+\hat{j}}
	+(\sigma+i\varepsilon\pi)_{x-\hat{j}}
	\right]
\ ,\label{Eq:meff}%
\end{align}
where $\sigma_x = \sum_{\vk,f(\vk)>0}
e^{i\bold{k}\cdot\bold{x}}\sigma_\vkt$
and
$\pi_x = \sum_{\vk,f(\vk)>0} (-1)^\tau e^{i\bold{k}\cdot\bold{x}}\pi_\vkt$.
Note that 
$\sigma_\vkmt=\sigma_\vkt^*$ and $\pi_\vkmt=\pi_\vkt^*$, 
since $\pm\bold{k}$ terms in \equref{Eq:SsFT}
are bosonized at a time.
At small $\vk$, they  are also regarded as the chiral ($\sigma$) fields and the Nambu-Goldstone
 ($\pi$) fields, respectively,
since the staggered fermion identifies the spin and flavor by specifying space-time position.
The sign factor, $\varepsilon_x=(-1)^{x_0+x_1+x_2+x_3}$, corresponds to
$\gamma_5\otimes {}^t\gamma_5$ in the spinor-taste space
in the continuum limit.

In the third step,
by integrating over the Grassmann and temporal link ($U_0$) variables
\cite{Faldt,BilicDeme,Bilic}, 
the partition function and the effective action are reduced to
\begin{align}
{\cal Z}_\mathrm{AF}%
=& \int \Fint[\sigma_{\vkt}, \pi_{\vkt}]~e^{-S_\mathrm{eff}^\mathrm{AF}}
\ , \\
S_\mathrm{eff}^\mathrm{AF}
=&\sum_{\vkt, f(\vk)>0}
  \frac{L^3f(\vk)}{4N_c}
  \left[\left|\sigma_\vkt\right|^2+\left|\pi_\vkt\right|^2\right]
-\sum_{\bold{x}} \log\RX
\ ,\label{Eq:SeffAF}\\
\RX
=& X_{N_\tau}(\bold{x})^3-2X_{N_\tau}(\bold{x})
	+2\cosh(\Nc \mu/T)
\end{align}
where 
$\Fint \left[ \sigma_{\vkt}, \pi_{\vkt} \right]=\prod_{\vkt,f(\vk)>0} d \sigma_{\vkt} d\sigma_{\vkt}^{\ast} d\pi_{\vkt} d\pi_{\vkt}^{\ast}$ and $T=\gamma^2 / \Nt$.
In the last line, we use a recursion formula to obtain $X_{N_\tau}(\bold{x})$~\cite{Faldt,Bilic,BilicDeme}.
In the cases where $m_{x=(\bold{x},\tau)}$ is static, 
we find
$X_{N_\tau}=2\cosh(N_\tau\ \mathrm{arcsinh}\ (m_x/\gamma))$.
It should be noted that the action represents the confinement phase,
since the baryonic chemical potential appears as $\cosh(N_c \mu/T)$,
which can be understood as baryonic contribution \cite{SC-Pol},  similarly to the PNJL
model with vanishing Polyakov loop \cite{Pol}.

We can now perform the Monte-Carlo integral
over the auxiliary fields $(\sigma_\vkt, \pi_\vkt)$
using the effective action $S_\mathrm{eff}^\mathrm{AF}$.
We generate Monte-Carlo 
configurations based on the phase quenched action 
at finite $\mu$ and $T$ and calculate observables
by the reweighting method;
\begin{align}
\langle \mathcal{O} \rangle 
= \frac{\langle \mathcal{O}\,\exp(i\theta) \rangle_\mathrm{pq}}
{\langle \exp(i\theta) \rangle_\mathrm{pq}}
\ ,
\end{align}
where $\theta=-\mathrm{Im}(S_\mathrm{eff}^\mathrm{AF})$
and $\langle \cdots \rangle_\mathrm{pq}$ denotes the phase quenched average.
Through this AFMC method,
we can numerically take account of fluctuation effects.

It should be noted that we have a sign problem that stems from the bosonization procedure,
but it is not severe and we can
investigate the QCD phase diagram as done in Ref.~\cite{Ichihara:2014ova}.
The reason for the milder sign problem may be understood as follows.
First, the auxiliary field effective action is obtained by
integrating out the link variables,
then it contains only the color singlet field.
Color singlet states are, in general, closer to the energy eigenstates
than colored states, and we expect smaller phases in the path integral.
Next, there is no sign problem in the mean field approximation,
where $\sigma_x$ is assumed to be constant and $\pi_x$ fields are neglected,
then the sign problem is not severe with small $|\bm{k}|$.
In the case where
the long wave physics dominates,
only small $\vert\bold{k}\vert$ auxiliary fields are relevant,
thus 
the $\pi$ field %
can be regarded as almost constant.
As seen
in \equref{Eq:meff}, the complex phase of one site
has an opposite sign 
to that of the
 nearest neighbor sites. Therefore, we could expect that the complex phase 
on one site
coming from the bosonization is canceled out 
by the nearest neighbor site contributions
as long as we study the 
long wave phenomena.
As a result, we have a milder sign problem
and can directly generate MC configurations at finite $\mu$ and $T$
as long as the lattice size is not very large. 
In actual calculations, high $\vert\bold{k}\vert$ modes are not 
negligible, and the average phase factor $\langle \exp(i\theta) \rangle_\mathrm{pq}$
is suppressed;
\footnote{The average phase factor indicates the severity of weight cancellation.
We have no weight cancellation when $\langle \exp(i\theta) \rangle_\mathrm{pq} = 1$.}
$\langle \exp(i\theta) \rangle_\mathrm{pq} \gtrsim 0.9$ and 0.4
 for $\mu/ T \lesssim 0.8$
 on the $4^3\times4$ and $8^3\times8$ lattices around the phase boundary, respectively~\cite{Ichihara:2014ova}.

\begin{figure}[!t]
 \begin{minipage}{0.5\hsize}
  \begin{center}
   \includegraphics[width=55mm, angle=270]{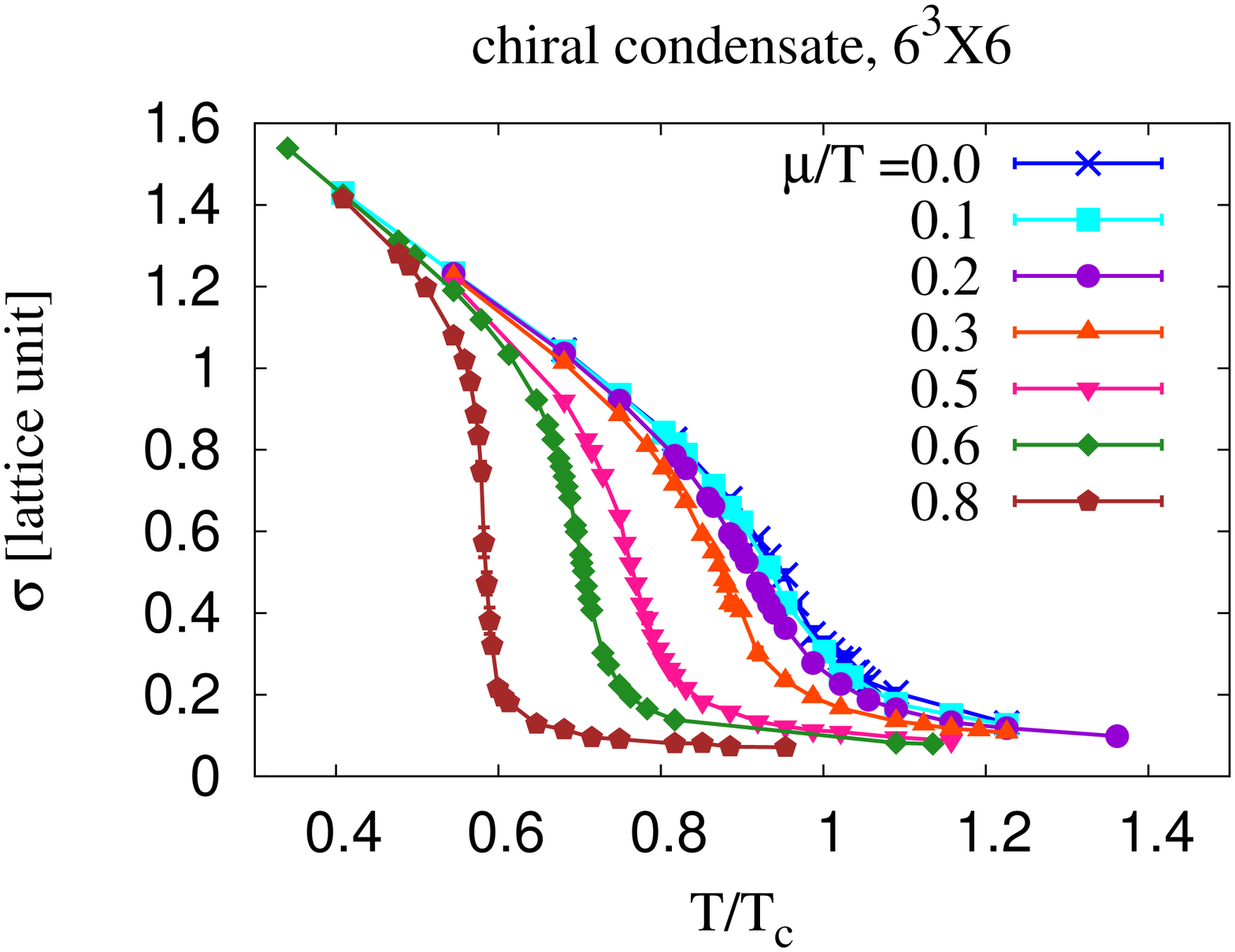}
  \end{center}
 \end{minipage}
 \begin{minipage}{0.5\hsize}
  \begin{center}
   \includegraphics[width=55mm, angle=270]{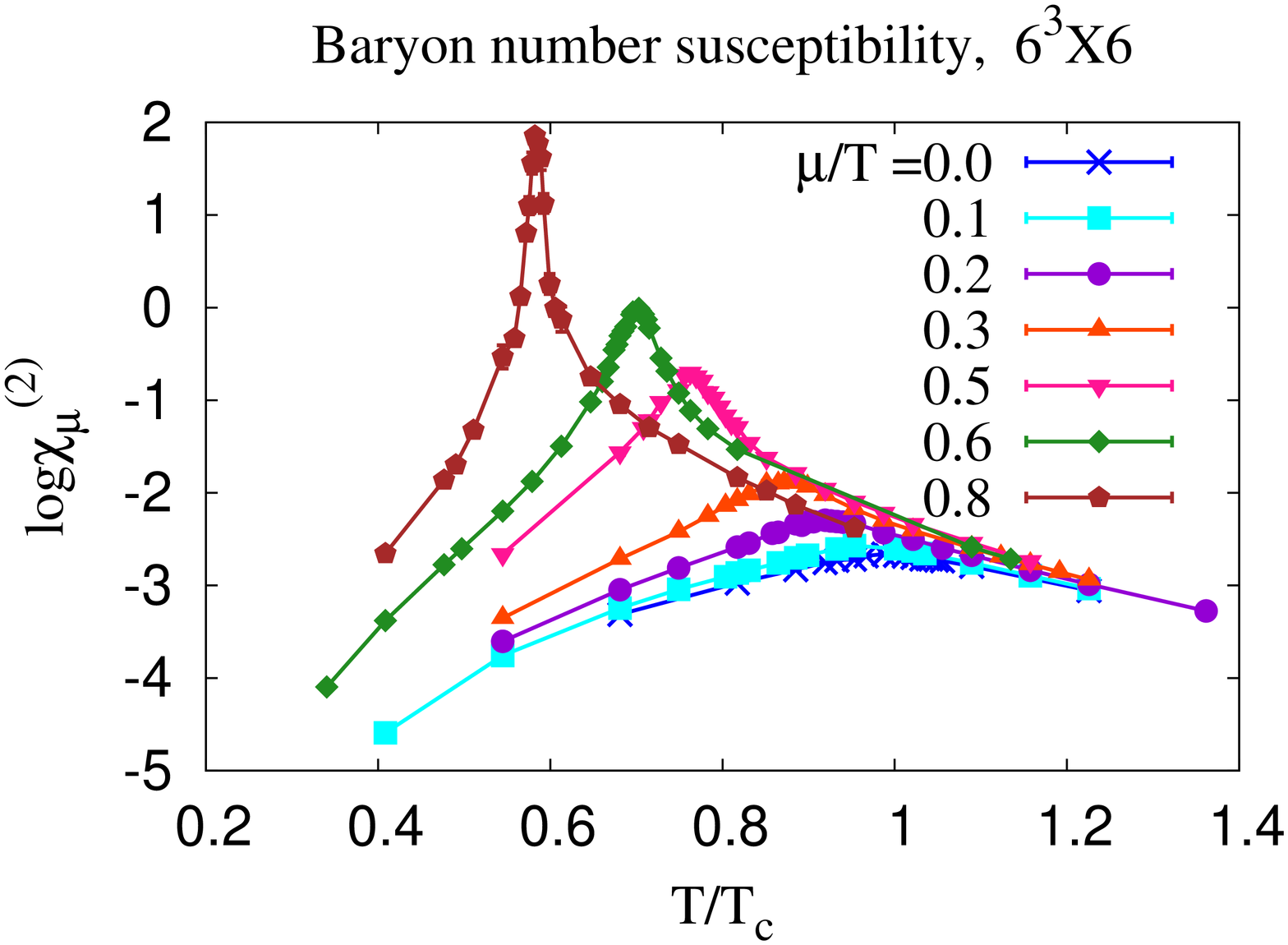}
  \end{center}
  \end{minipage}
   \caption{Chiral condensate and logarithmic scale of baryon
    number susceptibility on a $6^3\times 6$ lattice for various $\mu /T$ lines.
   The cross mark, square, circle, triangle, lower triangle, diamond, and pentagon denote
   $\mu / T = 0.0, 0.1, 0.2, 0.3, 0.5, 0.6$ and $0.8$, respectively.
     }
\label{Fig:orderpara}
\end{figure}

In \figref{Fig:orderpara},
we show the chiral condensate ($\sigma = \expv{\sum_{\tau} \sigma_{\bm{k}=\bf{0},\tau}/\Nt }$) and the logarithm of the baryon number susceptibility ($\CUM2 = \partial^2 \log Z/ \partial (N_c\mu /T )^2 /(VT^3)$) in the chiral limit on a $6^3\times 6$ lattice 
as a function of $T/T_c$,
where $T_c(\simeq 1.46862)$ denotes critical temperature at $\mu=0$
on a $6^3 \times 6$ lattice defined by the peak position of chiral susceptibility ($\chi_\sigma = \partial^2 \log Z / \partial m_0^2 / (L^3 \Nt)$)~\cite{Ichihara:2014ova}.
We can see the chiral phase transition behavior 
from the decrease of the chiral condensate with increasing $T$.
The decrease of the chiral condensate becomes steeper with increasing $\mu / T$.
The baryon number susceptibility 
has a sharp peak at high $\mu /T$.
These behaviors suggest the influence of the singularity at the tricritical point.
It should be noted that the baryon number susceptibility at zero chemical potential 
decreases at high $T$.
This trend is not consistent with standard lattice QCD results
and would be an artifact of the truncation of the action in \equref{Eq:Seff}, 
so 
we only focus on the vicinity of the phase transition.

\subsection{Some remarks on the universality class} 
The staggered fermion has a remnant chiral symmetry in the chiral limit
and has O(2) symmetry as long as the lattice spacing is 
coarse.
We introduce the auxiliary fields to maintain the
original symmetry, O(2). 
The auxiliary field action $\Seff^\mathrm{AF}$ in \equref{Eq:SeffAF}
is invariant under the chiral transformation,
\begin{align}
\begin{pmatrix}
\sigma_k \\ \pi_k
\end{pmatrix}
\to
\begin{pmatrix}
\sigma_k' \\ \pi_k'
\end{pmatrix}
=
\begin{pmatrix}
\cos\Upsilon & -\sin\Upsilon \\
\sin\Upsilon & \cos\Upsilon
\end{pmatrix}
\begin{pmatrix}
\sigma_k \\ \pi_k
\end{pmatrix}
\ ,
\label{Eq:chitransAF}
\end{align}
\com{which corresponds to the transformation of}
the staggered fermion fields
\begin{align}
\chi_x \to \chi_x' = e^{i\varepsilon_x \Upsilon/2} \chi_x
\ ,\quad
\chibar_x \to \chibar_x' = e^{i\varepsilon_x \Upsilon/2} \chibar_x\ .
\end{align}
In the framework of the MDP in the strong coupling limit, the critical
exponent was studied  and that values are consistent with
O(2)~\cite{MDP-CE}. 
It should be noted that we need careful discussions about the difference in symmetry
since the critical exponents are different
in $\mathrm{O}(2)$, $\mathrm{O}(4)$, and Z(2).
We will mention the relation between the O($N$) scaling function and critical behavior in \secref{Sec:ON}.

\section{Net-baryon number fluctuations in the strong coupling limit}
\label{sec:results}

\subsection{Net-baryon number cumulants} \label{sec:Cumu}

In this section, we present results on the higher-order 
cumulant ratios of the net-baryon number.
The $n$-th order cumulant of the net-baryon number
in the grand canonical ensemble is given by 
\begin{equation}
 \chi_\mu^{(n)} = \frac{1}{VT^3} \frac{\partial^n \log Z}{\partial \hat{\mu}^n } 
  \ , \
  \hat{\mu} = \Nc \mu / T
  \ . \label{eq:cumu}
\end{equation}
To discuss the effects of the phase transition on the net-baryon number
fluctuations, it is convenient to take the ratio of a higher-order
cumulant to the second order one \cite{CumuRatio,CumuRatio-1}
since the volume factor, $V$, in \equref{eq:cumu} cancels. 
We show the results of the following cumulant ratios, the normalized skewness and kurtosis,
\begin{align}
 S \sigma &= \CUM3 / \CUM2\ , \quad
\kappa \sigma^2  = \CUM4 / \CUM2 
.
 \end{align}
The skewness $S$ and kurtosis $\kappa$ probe the asymmetry with respect
to the mean and peakedness 
of the underlying probability distribution, respectively.
The normalization by $\CUM2$ implies that one sets a reference,
since $\kappa\sigma^2 =1$ for the Skellam distribution which describes
the distribution of the difference $n_1-n_2$, where $n_1$ and $n_2$
follow the Poisson distribution. 
Then it corresponds to
the free hadron gas with Boltzmann approximation. 

The detailed method  for the calculation of the cumulants are summarized  in \secref{Sec:deriva}.
As discussed in \secref{Sec:deriva}, observables have an imaginary part 
due to the high $|\bf{k}|$ modes of $\pi_{\vkt}$,
so we take the real part of the observables and error bars in this article.
The absolute mean value of maximum imaginary part for the normalized kurtosis and skewness are about 70 and 2 for $\mu/T = 0.8$ and 0.5 and 0.04 for $\mu / T = 0.2$, respectively,
which are smaller than
the peak heights and valley depth of the real part
as seen in Figs.~\ref{Fig:sk} and \ref{Fig:kur}. 
It should be noted that
the high temperature behavior of the cumulants
would bare the artifacts of the present treatment.
Figures~\ref{Fig:sk} and \ref{Fig:kur} indicate that
$S\sigma$ stays negative in the large chemical potential region
and $\kappa\sigma^2$ can also take small negative values
depending on $\mu/T$
at high $T$. 
For a free ideal baryon 
gas
we expect
$S\sigma \to 6\hat{\mu} /(3\hat{\mu}^2 + \pi^2)$ and 
$\kappa\sigma^2 \to 6/( 3\hat{\mu}^2 + \pi^2)$
 at high temperature~\cite{CumuRatio,Stokic:2008jh,Allton:2005gk}.
These differences at high temperature may be due
to the truncation in the $1/d$ expansion.
Spatial baryon hopping term is missing
in the leading order in the $1/d$ expansion,
then %
the fermion momentum integral %
generates only a volume factor.
As a result, pressure is proportional to $T$
rather than $T^4$ in the Stefan-Boltzmann behavior at high temperature.
Since
we are interested in the transition region,
where we expect that the long wave mesonic fluctuations dominate,
we focus on the cumulant ratios around the phase boundary
in the later discussion.

\subsection{$T$ and $\mu$ dependence}

\begin{figure}[tbhp]
 \begin{minipage}{0.5\hsize}
  \begin{center}
   \includegraphics[width=55mm, angle=270]{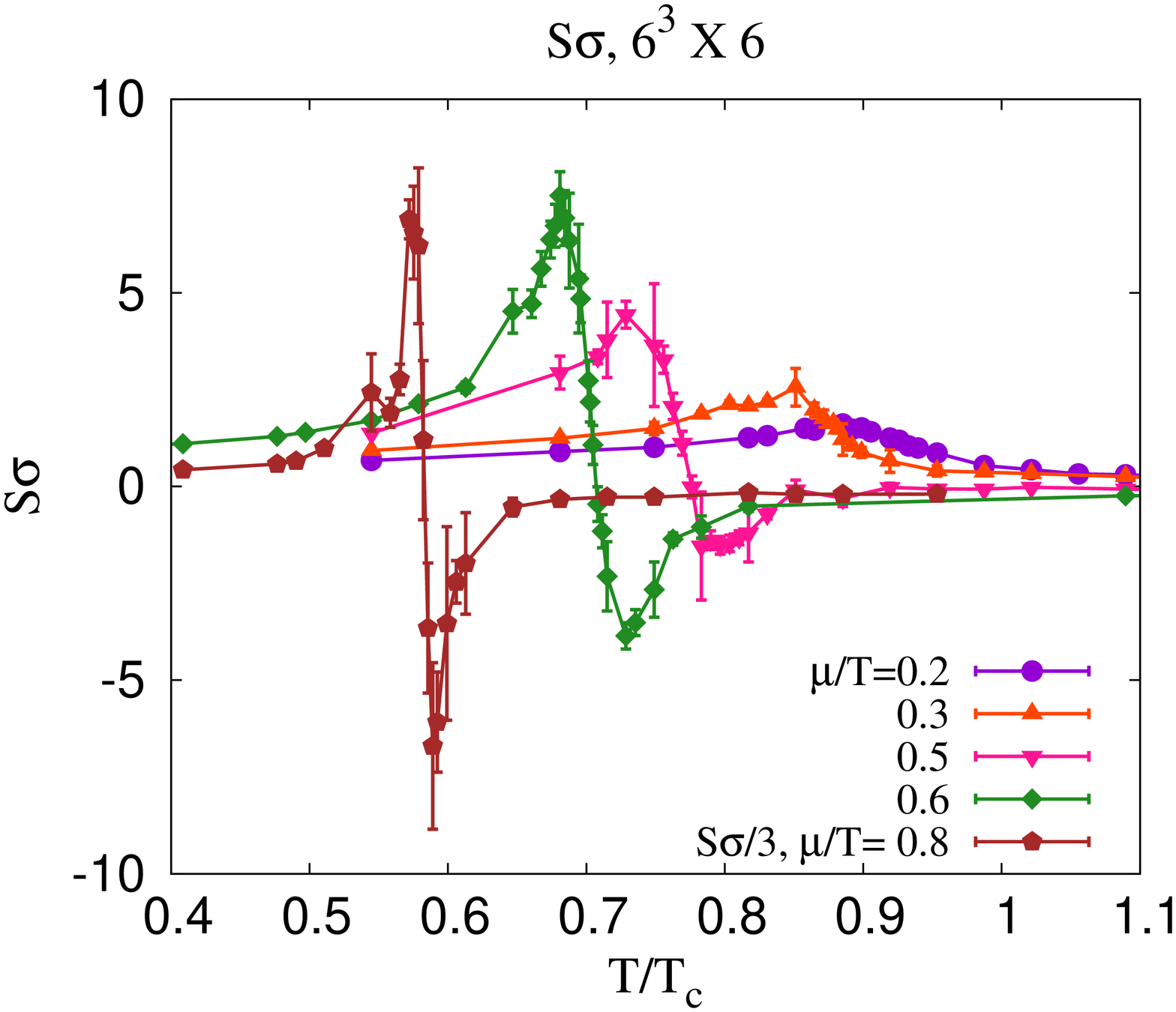}
  \end{center}
  \label{fig:one}
 \end{minipage}
 \begin{minipage}{0.5\hsize}
  \begin{center}
   \includegraphics[width=55mm, angle=270]{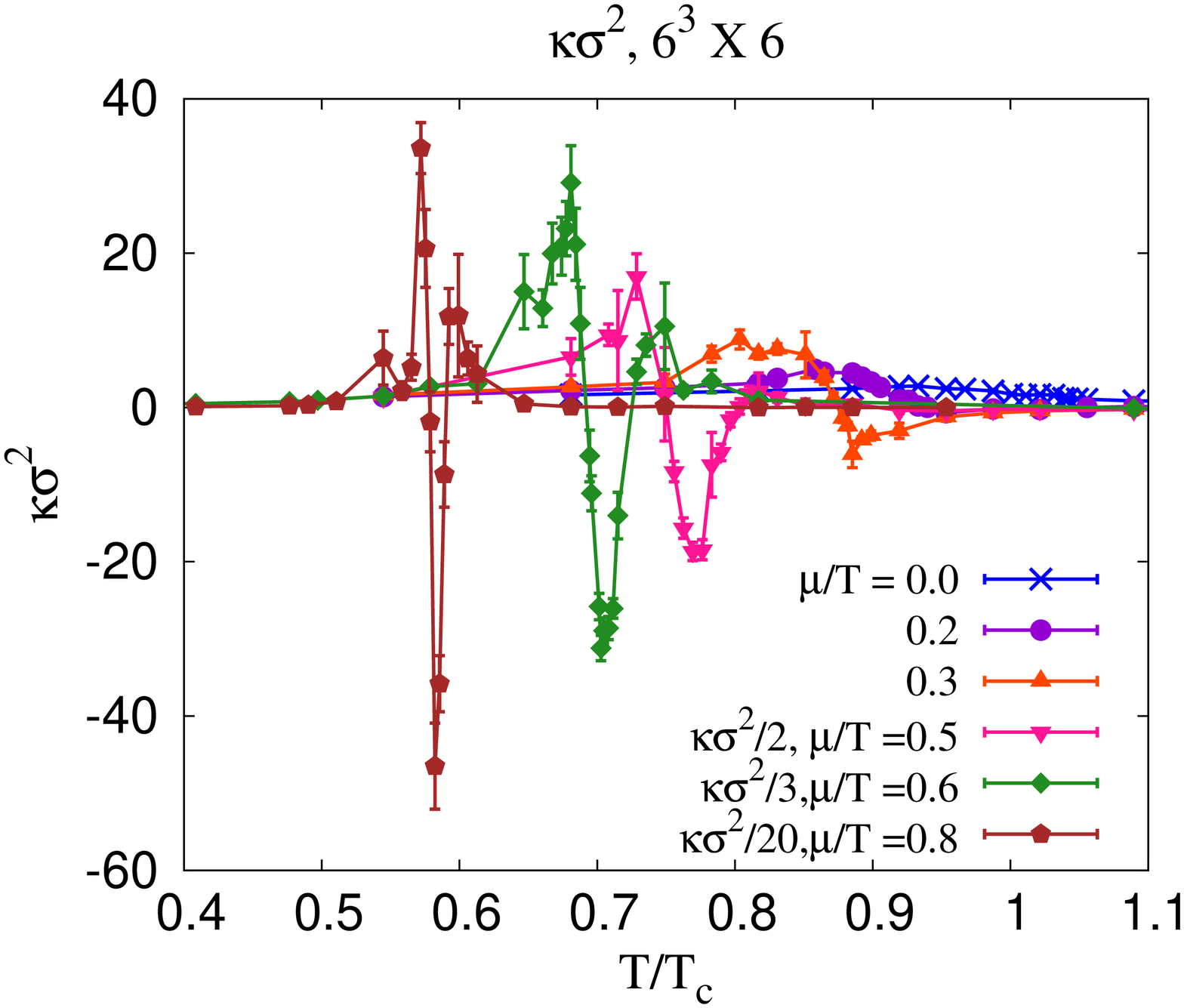}
  \end{center}
  \label{fig:two}
 \end{minipage}\\
 
 \caption{Normalized skewness (left panel) and kurtosis (right panel) on the
 various $\mu /T$ lines on a $6^3 \times 6$ lattice  The cross mark,
 circle, triangle, lower triangle, diamond, and pentagon indicate 
 $\mu / T =0.0,\ 0.2,\ 0.3,\ 0.5,\ 0.6,$ and $0.8$ line, respectively.
 As for the skewness on the $\mu / T =0.8$ line,
 we rescale results by factor 3.
 For the kurtosis on the $\mu / T =0.5,0.6,0.8$ lines,
 we rescale results by factor 2, 3, 20, respectively. 
 }
 \label{Fig:L6N6}
\end{figure}

In \figref{Fig:L6N6}, we show the results of $S\sigma$ and 
$\kappa\sigma^2$ on a $6^4$ lattice %
at several
$\mu /T$ %
as a function of $T/T_c$. One sees the strong dependence of 
these higher-order cumulants on both temperature and chemical potential.
In the low chemical potential region, both $S\sigma$ and
$\kappa\sigma^2$ are positive and have a broad peak.

As $\mu / T$ increases, characteristic structures emerge.
$S\sigma$ increases with $T$, then exhibits a strong positive peak
followed by sudden decrease to large negative value. Further increase
of temperature leads to a constant small value. 
The negative region of the skewness,
indicating the critical behavior \cite{CModel-3rd},
starts to appear
between $\mu /T = 0.3$ and $\mu /T = 0.5$.
$\kappa\sigma^2$ shows similarly moderate increase
with temperature and a sharp positive peak. 
Then it turns to be negative
with a sharp %
valley but becomes positive again
with a somewhat milder peak.
Such an oscillatory behavior appears in $\mu /T > 0.2$.
Both cumulant ratios have one negative %
valley,
and the skewness (kurtosis) %
has one (two) positive %
 peak(s)
at high chemical potential.
Comparing with the phase boundary \cite{Ichihara:2014ova},
one finds these behaviors appear 
around the phase boundary (see below).
In the following subsections, we discuss the behaviors in more
details by looking at lattice size dependence.

\subsection{Lattice size dependence} \label{sec:Sk}

\begin{figure}[!t]
 \begin{minipage}{0.5\hsize}
  \begin{center}
   \includegraphics[width=55mm, angle=270]{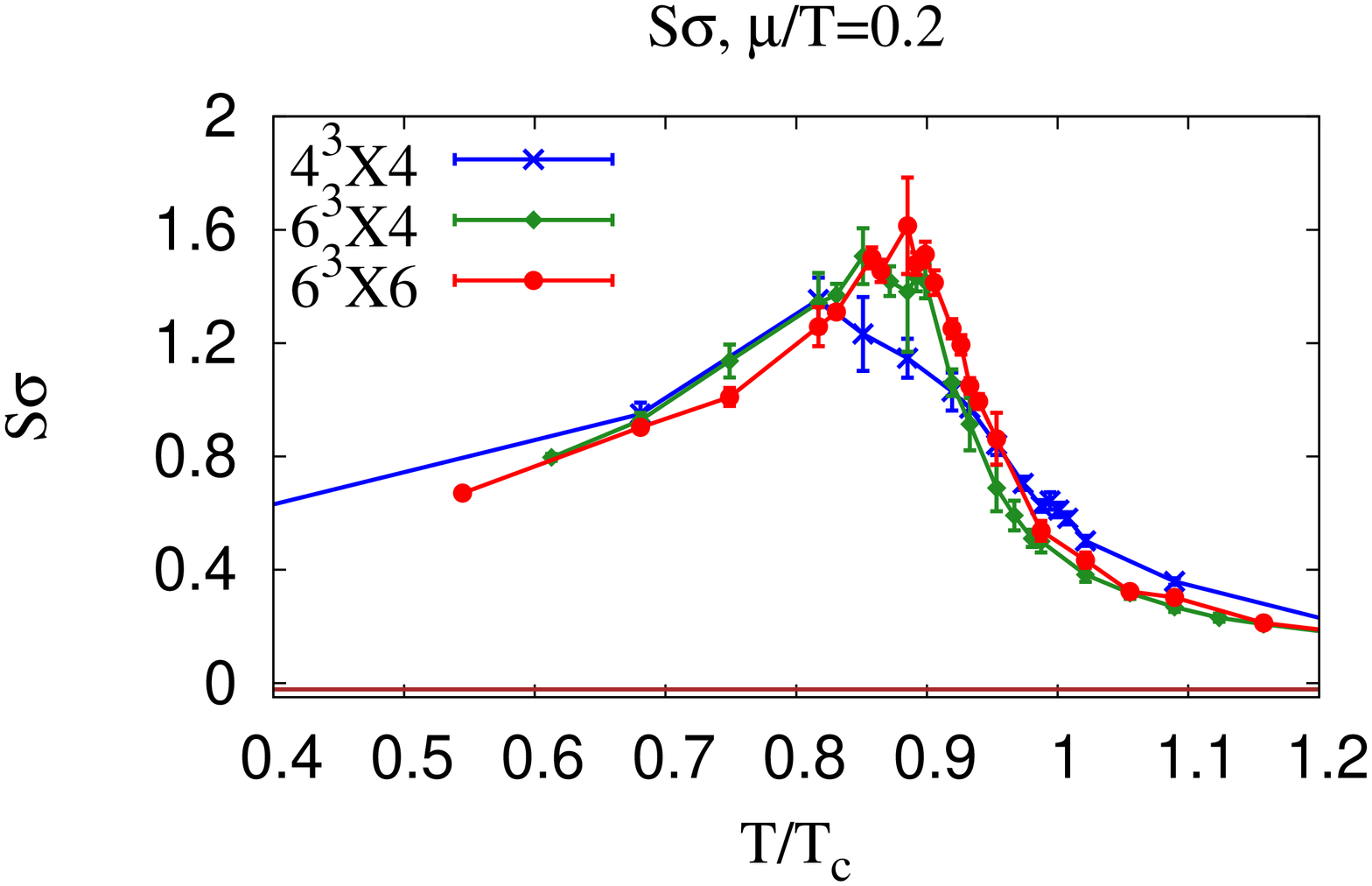}
  \end{center}
 \end{minipage}
 \begin{minipage}{0.5\hsize}
  \begin{center}
   \includegraphics[width=55mm, angle=270]{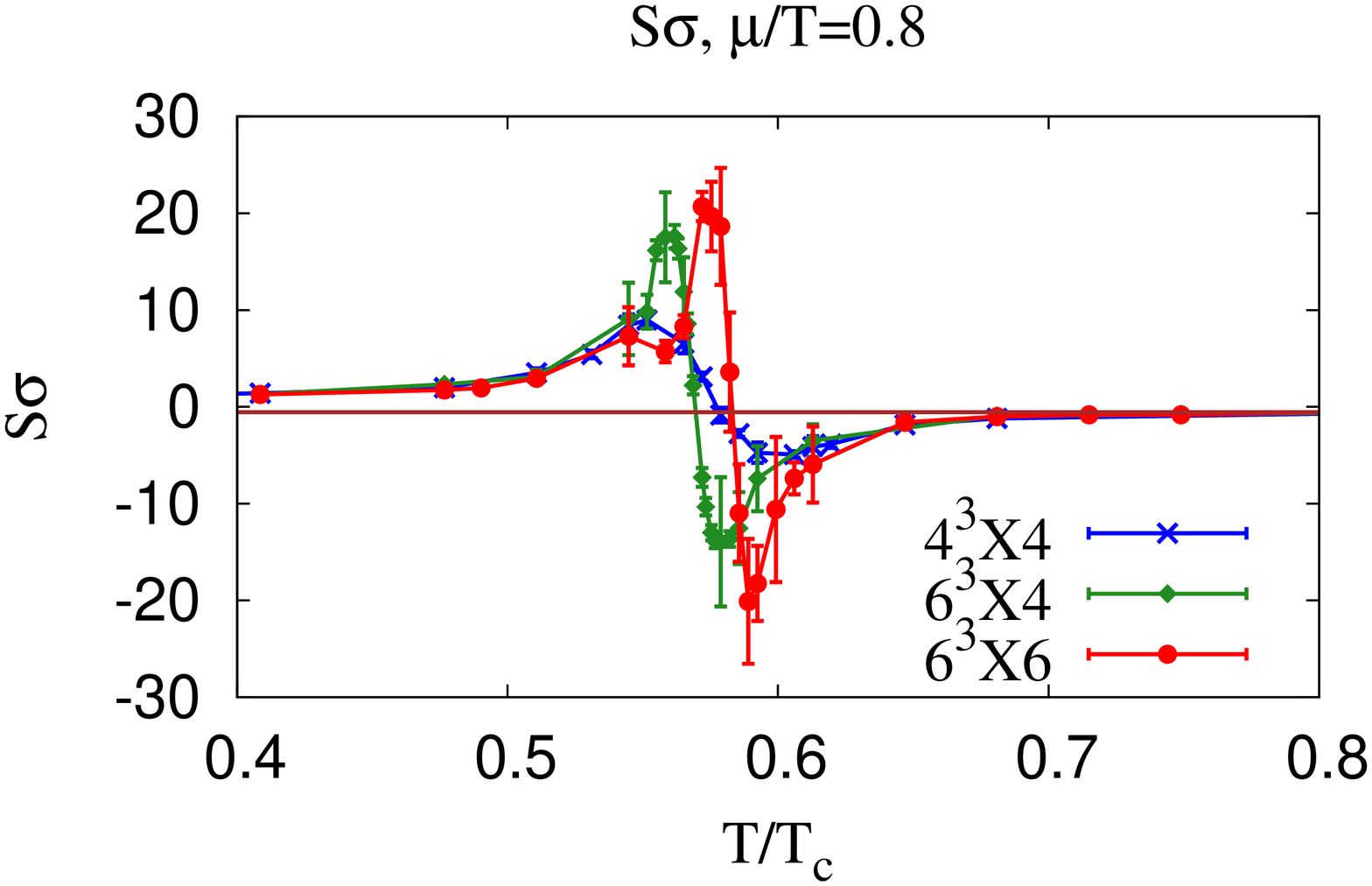}
  \end{center}
 \end{minipage}
  \caption{Normalized skewness on the $\mu /T=0.2$ (left panel) and $\mu
 /T=0.8$ (right panel) line on $4^3 \times 4$ (the cross mark), $6^3
 \times 4$ (the diamond mark), and $6^3\times 6$ (the circle mark)
 lattices. A horizontal line around zero denotes the mean filed value at
 high temperature (see text). 
 ($S \sigma \simeq -0.0220$ for $\mu /T =0.2$ and $S\sigma \simeq -0.5442$ for $\mu /T =0.8$.)
 We can find the negative region in the high chemical potential region.
  }
  \label{Fig:sk}
\end{figure}

\begin{figure}[!t]
 \begin{minipage}{0.5\hsize}
  \begin{center}
   \includegraphics[width=55mm, angle=270]{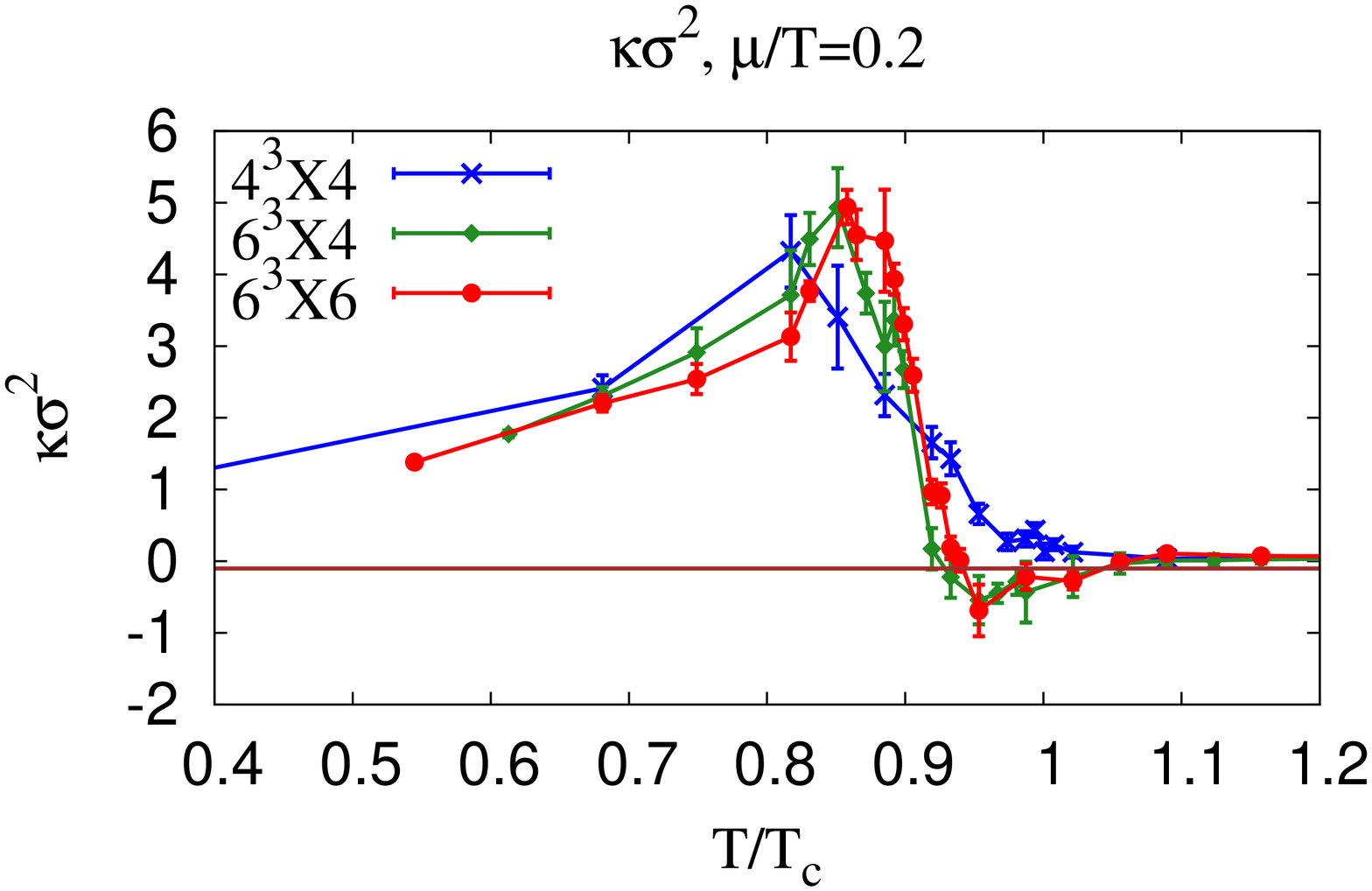}
  \end{center}
 \end{minipage}
 \begin{minipage}{0.5\hsize}
  \begin{center}
   \includegraphics[width=55mm, angle=270]{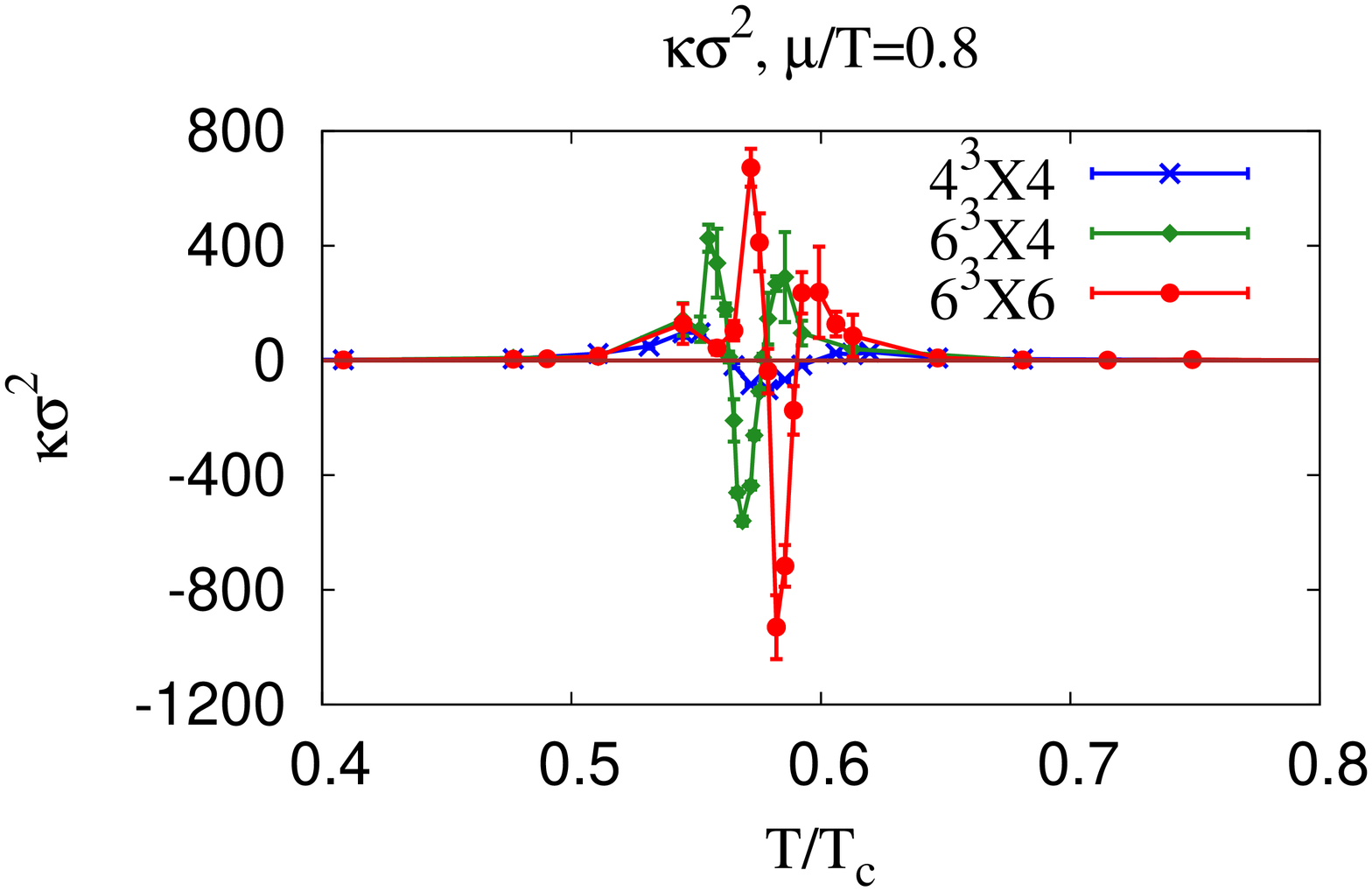}
  \end{center}
 \end{minipage}
   \caption{Normalized kurtosis on the $\mu /T=0.2$ (left panel) and $\mu /T=0.8$ (right panel) line on $4^3 \times 4$ (the cross mark), $6^3 \times 4$ (the diamond mark), and $6^3\times 6$ (the circle mark) lattices.
     The thin horizontal line around zero denotes the mean field value at high
 temperature (see text).
 ($\kappa \sigma^2 \simeq -0.1033$ for $\mu /T =0.2$ and $\kappa \sigma^2 \simeq -0.0251$ for $\mu /T =0.8$.)
 }
\label{Fig:kur}
\end{figure}

In Figs.~\ref{Fig:sk} and \ref{Fig:kur},
we show the results 
of the normalized skewness $S\sigma$ 
and the normalized kurtosis $\kappa\sigma^2$,
respectively,
on the fixed $\mu /T$ lines,
$\mu/T=0.2$ (left panel) and $\mu/T=0.8$ (right panel), on
$4^4, 6^3\times 4$ and $6^4$ lattices.
Horizontal lines show %
the mean field values at high temperature, 
where the chiral condensate vanishes.
We will use these mean field values in \secref{sec:kur-inPD}
to remove the artifact 
as discussed in \secref{sec:Cumu}
%

Lattice size dependence of $S\sigma$ is
moderate at low $\mu/T$ and prominent at large $\mu/T$.
From Fig.~\ref{Fig:sk}, one sees that the qualitative structure of the
temperature dependence of $S\sigma$ does not strongly depend on the lattice size.
For $\mu/T=0.2$, %
the skewness is positive around the phase transition region
and has a peak.
While the peak height does not show statistically significant dependence
on the lattice size, one %
finds the temperature dependence
becomes slightly stronger in the largest one, $6^4$.
The lattice size dependence becomes prominent at large chemical
potential $\mu/T=0.8$.
The skewness has %
one positive peak
and %
one negative valley at
$\mu / T = 0.8$.
The widths of the  %
peak and the valley
become narrower on a $6^4$ lattice.

Similar lattice size dependence is found for $\kappa\sigma^2$.
In the low chemical potential region ($\mu/T=0.2$),
the smallest lattice size $4^4$ does not exhibit the characteristic
structure seen in %
larger lattice cases,
but one sees a moderate
decrease around $T/T_c\simeq 0.9$ following a peak %
at lower $T$.
The broad negative %
valley
appears %
on $6^3\times4$ and $6^4$ lattices,
implying that
one %
should not
take too small lattice size to see the influence of the critical fluctuations. 
We find that the oscillatory behavior with two positive peaks
and one negative valley
and its lattice size dependence becomes prominent for $\mu/T = 0.8$.
Again, as lattice size becomes larger,
the peak height tends to be larger value
and the width of the oscillatory region becomes narrower.

\subsection{Relation to \rm{O($N$)} scaling} \label{Sec:ON}
The behavior of the higher-order net-baryon number cumulants around
chiral phase transition are expected to be described by the property of
the O($N$) scaling function \cite{O2-CE-alpha,O4-CE-alpha,SF-4,O2-O4-study,SF-4-CM}.
According to Ref.~\cite{SF-4-CM}, 
the first divergent cumulant at $\mu=0$ in the thermodynamic limit
is the sixth order one $\CUM6$ for $N=2$ and $4$.
This result comes from the fact that the critical exponent of the
specific heat $\alpha$ is negative in these cases, $\alpha\simeq -0.0147$ in
O(2) \cite{O2-CE-alpha} and $\alpha\simeq -0.21$ in O(4) \cite{O4-CE-alpha}. For positive $\alpha$, the
divergence appears at the fourth order cumulants while it has a cusp for
$\alpha < 0$ \cite{FRG-HM,Morita_prob}.

For a nonzero $\mu$, the first divergent cumulant is the third order
one thus $S\sigma$ diverges in the thermodynamic limit for 3d O(2) and O(4)
universality class. Once the singular part is smeared by the explicit
breaking (nonzero current quark mass) or finite volume effects
\footnote{
Recalling the behavior of the order parameters becomes moderate not only by finite mass
but also finite size effects,
we could guess that the singular part of the cumulants is also masked by the finite size effects.
By using models, we can find the characteristic behavior of physical quantities becomes smoother when the system is finite \cite{ModelStudy-Volume-effect}. 
In the framework of 3d Ising model, finite size results are studied in Ref.~\cite{Chen:2010ej}
and the peak heights of $S\sigma$ and $\kappa \sigma^2$ increase with larger volume 
due to the correlation length $\xi$. %
The relations between higher-order net-baryon number cumulants and the correlation length in 3d O(4) spin model are pointed out in Ref.~\cite{chiral-model-and-scaling}.
}, 
the leading singular contribution is suppressed by
the small multiplicative
factor $-(2\kappa_q) (\mu/T)^n$ where $\kappa_q$ denotes the curvature of
the phase boundary near $\mu=0$ in the chiral limit \cite{SF-4-CM}.
Then, whether one sees the effects of critical fluctuations in the
higher-order cumulants or not depends on the value of $\mu/T$ and
magnitude of the regular part of the free energy density and its derivatives.
For instance, $S\sigma$ positively (negatively) diverges 
in the thermodynamic limit
when approached from below (above) the phase transition.
The negative divergent contribution is replaced by a large negative value in the
presence of the explicit breaking or finite volume effects.
Then, the appearance of the negative $S\sigma$  depends on  whether the
singular part overcomes the regular part.
Similar argument applies to $\kappa\sigma^2$, and the
absence of the negative
region in %
$\kappa\sigma^2$
in the smallest lattice %
(\figref{Fig:kur})
is a consequence of the finite volume effects.
Since 
$\kappa\sigma^2$
exhibits the oscillation
around the phase transition along constant $\mu/T$ %
including the narrow negative region between the two positive peaks,
one expects 
$\kappa \sigma^2$ positively diverges both from below and above the phase transition
and negative region would disappear in the thermodynamic limit. 
The lattice size dependence of the negative $\kappa \sigma^2$ region
meets this expectation.
This is in contrast to the case of the finite quark mass.
With finite quark mass,
one finds the negative kurtosis region even in the thermodynamic limit as a remnant of the divergence 
in the chiral limit \cite{SF-4-CM,FRG-HM}.

\subsection{Kurtosis in the QCD phase diagram in the strong coupling limit}
\label{sec:kur-inPD}
\begin{figure}[tbhp]
\centering
   \includegraphics[width=65mm, angle=270]{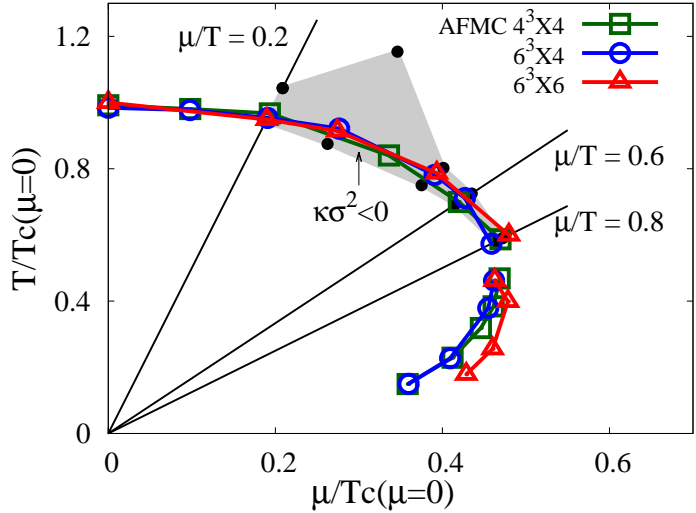}
\vspace{5mm}
  \caption{ 
  The negative kurtosis region in the QCD
  phase diagram in the chiral and strong coupling limit. 
 The shaded area denotes the region where the $\kappa \sigma^2$ is 
  both 
  negative 
  and 
  smaller than the mean field value
  on a $6^4$ lattice (more detail explanations are in the text and footnote).
  The vertical axis is the temperature parametrized as $T=\gamma^2/\Nt$
  and
  the horizontal axis is the quark chemical potential.
  Both $T$ and $\mu$ are normalized by using $T_c$, the critical temperature at $\mu =0$ on a $6^4$ lattice.
  The phase boundaries 
  on $4^3 \times 4, 6^3 \times 4$, and $6^3\times 6$ lattices are determined by the peak position of the chiral susceptibility
  ($\chi_\sigma = \partial^2 \log Z / \partial m_0^2 / (L^3 \Nt)$)
  for $\mu / T \le 0.8$
  and the effective potential analysis
  defined by 
  $F_{\mathrm{eff}} = \expv{S_{\mathrm{eff}}}/(\Nt L^3)$ for $\mu / T \ge 1.0$~\cite{Ichihara:2014ova}.
  A gap is the expected boundary 
  between the would-be 1st order and the would-be 2nd order %
   transition.
   }
  \label{Fig:kurPB}
\end{figure}

In \figref{Fig:kurPB},
we show the negative kurtosis region %
in the chiral limit $(m_0 \rightarrow 0)$ on a $6^4$ lattice.
We also show the
 phase boundary %
 determined by the peak position 
of the chiral susceptibility
($\chi_\sigma = \partial^2 \log Z / \partial m_0^2 / (L^3 \Nt)$)
for $\mu / T \le 0.8$
  and by the cross point of the effective potential 
  defined by $F_{\mathrm{eff}} = \expv{S_{\mathrm{eff}}}/(\Nt L^3)$
  for $\mu / T \ge 1.0$.
The transition 
would be the second order for $\mu / T \le 0.8$. 
While the numerical finite size scaling 
analysis cannot rule out crossover due to the statistics~\cite{Ichihara:2014ova}, 
we expect
the second order transition from symmetry arguments~\cite{PW84}.
We 
guess that the tricritical 
point exists at $\mu/T > 0.8$.

We here define the negative kurtosis region %
where $\kappa\sigma^2$ is smaller
than the mean field results at high $T$,
in order to reduce the artifact at high $T$ as discussed in %
\secref{sec:Cumu}.
It should be noted that we do not take account of error bars to define the region here.
The negative kurtosis region appears from about $\mu /T =0.2$,
and the region is the largest %
at $\mu /T = 0.3$
\footnote{
Along the $\mu / T =0.3$ line at high $T$, 
the mean field value and AFMC results are so close to each other 
and 
it seems that the AFMC results oscillate around the MF result.
Thus we define the boundary of the negative kurtosis region as the first intersection of the AFMC and MF results.
}. 
By comparison, the kurtosis is positive %
at $\mu / T < 0.2$.
This might be due to
the suppression of the singular contribution %
by the factor $(\mu/T)^n$ and the regular part dominates 
over the singular part \cite{SF-4-CM}. 

The negative $\kappa\sigma^2$ region %
almost coincides with the phase boundary. %
As discussed in %
Secs.~\ref{sec:Sk} and \ref{Sec:ON},
the region will shrink to the phase boundary 
and finally vanish in the thermodynamic limit.
Although the dependence of the scaling function is different
between the finite size effect and the symmetry breaking term,
one could expect that the negative region also appears 
with finite mass in the strong coupling limit.

We conclude that we find the negative skewness or kurtosis region in the
AFMC method in the chiral %
and strong coupling limit 
due to the finite volume effects as for the kurtosis,
as a consequence of the critical fluctuations around the phase boundary
incorporated through the AFMC method.

\section{Summary} \label{sec:summary}
We have
investigated the higher-order cumulant ratios of the net-baryon number,
$ S \sigma = \CUM3 / \CUM2 
 \ , \kappa \sigma^2  = \CUM4 / \CUM2$,
in the strong coupling $(g\rightarrow \infty)$
and the chiral limit ($m_0 \rightarrow 0$)
of QCD in the leading order of large dimensional expansions.
Mesonic fluctuation effects are taken into account by making use of 
the auxiliary field Monte-Carlo (AFMC) method.
We find the negative skewness and the kurtosis region
around the boundary 
of the chiral phase transition.
The skewness and kurtosis exhibit characteristic temperature dependences
influenced by the critical fluctuations.

The skewness is found to be negative near the phase boundary
for large $\mu/T$.
The oscillatory behavior around the phase boundary seems to be
consistent with the expectations from the finite volume effect such that
the positive (negative) peak at lower (higher) temperature diverges in
the thermodynamic limit \cite{SF-4-CM}.
Similarly, the kurtosis has one negative %
valley between two positive peaks for large $\mu/T$.
With increasing lattice size,
the %
negative valley is found to %
shrink,
as anticipated for the finite volume effect.
Thus, we expect two positive peaks diverge and the negative region disappears
in the thermodynamic limit \cite{SF-4-CM}.

One of the important next steps 
could be to study the effect of finite mass. %
This can be carried out by applying the AFMC method.
Another important step is to see the finite size effects \cite{FSS,ModelStudy-Volume-effect,Vol-effect-study}
and determine the negative kurtosis area with or without bare quark mass.
Whether one could have such region in lattice QCD %
is
a mandatory question.
As discussed above, it will depend on whether the smeared singular
contribution by finite volume and/or explicit symmetry breaking term can overwhelm
the regular contribution~\cite{SF-4-CM,ModelStudy-Volume-effect}, 
which corresponds to hadron resonance gas in
finite temperature QCD below the phase transition.
Since our present formulation ignores the 
spatial baryon hopping,
we would expect
that the regular contribution might be smaller than the realistic case
thus our results could serve as a lower limit for the value of chemical
potential where $\kappa \sigma^2 <0$. 

\section*{Acknowledgment}
The authors would like to thank Sinya Aoki, Frithjof Karsch, Swagato Mukherjee, Hiroshi Ohno, 
Philippe de Forcrand, Hideaki Iida, Yu Maezawa, Keitaro Nagata,  Shuntaro Sakai, Takahiro Sasaki, 
 and Wolfgang Unger
 for useful discussions.
Part of numerical computation in this work was carried out at the Yukawa Institute Computer Facility.
TI is supported by the Grants-in-Aid for JSPS Fellows (No.25-2059)
and would like to thank for RIKEN-BNL Brain Circulation Program. 
KM was supported by Polish Science Foundation (NCN),
under Maestro grant 2013/10/A/ST2/00106.
This work is supported in part by the Grants-in-Aid for Scientific Research
from JSPS
(Nos.
	23340067, 
	24340054, 
	24540271, 
	15K05079
),
by the Grants-in-Aid for Scientific Research on Innovative Areas from MEXT
(No. 2404: 24105001, 24105008), 
by the Yukawa International Program for Quark-hadron Sciences.

\appendix

 \section{Higher order derivatives of the baryon chemical potential}
 \label{Sec:deriva}
 
 In this appendix, we show the higher order derivatives with respect to dimensionless chemical potential $\hat{\mu} (= \Nc \mu / T)$.
 In the following, the action $S$ and the partition function $Z$ are denoted as 
 $S_{\mathrm{eff}}^{\mathrm{AF}}$ and ${\cal Z}_\mathrm{AF}$, respectively.
 Then, higher order derivatives are given as


\begin{align}
 \CUM1& = \frac{1}{VT^3} \frac{\partial (\log Z)}{\partial\hat{\mu}} 
 =  \frac{1}{VT^3} \left[ \frac{1}{Z} \int \mathcal{D} \Phi \left( -\frac{\partial S}{\partial \hat{\mu}} \right)  e^{-S}\right] =  \frac{1}{VT^3} \left< - \frac{\partial S}{\partial \hat{\mu}} \right>
  \ , \\
 \CUM2 &= \frac{1}{VT^3} \frac{\partial^2(\log Z)}{\partial\hat{\mu}^2} 
= \frac{1}{VT^3} \left[ \expv{   \left(  \frac{\partial S}{\partial \hatmu} - \expv{ \frac{\partial S}{\partial \hatmu} } \right)^2 }- \expv{ \frac{\partial^2 S}{\partial \hat{\mu} ^2}}  \right]
  \ , \\
 \CUM3 &= \frac{1}{VT^3} \frac{\partial^3(\log Z)}{\partial\hat{\mu}^3} 
 \non \\
 & = \frac{1}{VT^3} \left[ -  \expv{  \left(  \frac{\partial S}{\partial \hatmu} - \expv{ \frac{\partial S}{\partial \hatmu}} \right)^3 }  
 - \expv{ \frac{\partial^3 S}{\partial \hatmu^3} }
  + 3 \expv{ \left(  \frac{\partial S}{\partial \hatmu} - \expv{ \frac{\partial S}{\partial \hatmu}} \right) 
 \left(  \frac{\partial^2 S}{\partial \hatmu^2} - \expv{ \frac{\partial^2 S}{\partial \hatmu^2}}  \right)  }
 \right]
  \ , \\
  \CUM4 &= \frac{1}{VT^3} \frac{\partial^4(\log Z)}{\partial\hat{\mu}^4}
  \non \\
  & = \frac{1}{VT^3}
  \left[ 
  \expv{\left( \frac{\partial S}{\partial \hatmu} - \expv{\frac{\partial S}{\partial \hatmu}} \right)^4  }
  -3  \expv{\left( \frac{\partial S}{\partial \hatmu} -\expv{\frac{\partial S}{\partial \hatmu}} \right)^2 }^2   
  + 3 \expv{  \left ( \frac{\partial^2 S}{\partial \hatmu^2}  - \expv{\frac{\partial^2 S}{\partial \hatmu^2}} \right)^2 }
  \right. \non \\
  &\left. 
  -6 \expv{ \left\{ \left(\frac{\partial S}{\partial \hatmu} \right)^2  - \expv{\left( \frac{\partial S}{\partial \hatmu}\right)^2} \right\} \left( \frac{\partial^2 S}{\partial \hatmu^2} -\expv{\frac{\partial^2 S}{\partial \hatmu^2}} \right)} 
  \right. \non \\
&  \left.
  + 4 \expv{ \left( \frac{\partial S}{\partial \hatmu} - \expv{ \frac{\partial S}{\partial \hatmu}  } \right) 
  \left( \frac{\partial^3 S}{\partial \hatmu^3} - \expv{\frac{\partial^3 S}{\partial \hatmu^3} }   \right) }
  \right. \non \\
  & \left.
  + 12 \expv{\frac{\partial S}{\partial \hatmu}} 
  \expv{\left( \frac{\partial S}{\partial \hatmu} - \expv{\frac{\partial S}{\partial \hatmu}} \right)
   \left( \frac{\partial^2 S}{\partial \hatmu^2} - \expv{\frac{\partial^2 S}{\partial \hatmu^2}  } \right) }
   - \expv{\frac{\partial^4 S}{\partial \hatmu^4}}
  \right]
  \ ,
\end{align}
where $\hat{\mu} = \Nc \mu / T$.
The derivative\com{s} 
of the action 
with respect to $\hatmu$ are given as 
\begin{align}
\frac{\partial S}{\partial \hatmu} 
&= - \sum_{\bm{x}} \frac{1}{\RX} 2 \sinh \hatmu
\ , \\
\frac{\partial^2 S}{\partial \hatmu^2} 
& = \sum_{\bm{x}} \left[ \frac{1}{\RX^2}  (2\sinh \hatmu)^2 -\frac{1}{\RX} 2 \cosh \hatmu \right]
\ , \\
\frac{\partial^3 S}{\partial \hatmu^3} 
& = \sum_{\bm{x}} \left[ - \frac{2}{\RX^3}( 2 \sinh \hatmu)^3 + \frac{3}{\RX^2} (2\sinh \hatmu)(2\cosh \hatmu ) - \frac{1}{\RX} 2 \sinh \hatmu \right]
\ , \\
\frac{\partial^4 S}{\partial \hatmu^4}
& =\sum_{\bm{x}} \left[ \frac{6}{\RX^4} (2 \sinh \hatmu)^4 - \frac{12}{\RX^3} (2\sinh \hatmu)^2 2\cosh \hatmu \right.
\non \\
&\left. \hspace{1.5cm} + \frac{3}{\RX^2} (2\cosh \hatmu)^2 + \frac{4}{\RX^2} (2\sinh \hatmu)^2 - \frac{1}{\RX} 2 \cosh \hatmu  \right]
\ ,
\end{align}
where $S = \sum_{\bm{k},\tau, f > 0} \frac{L^3}{4\Nc} f(\bm{k}) \left[ |\sigma_{\bm{k},\tau}|^2 +|\pi_{\bm{k},\tau}|^2 \right] - \sum_{\bm{x}} \log \RX $
and $\RX = X_{N}^{3} (\bm{x})- 2 X_N(\bm{x}) + 2 \cosh\hatmu $.

In this analysis, we have a complex phase coming from the fermion
determinant, so the observables also take a complex value
even if the observables are real in essence. %
Invoking the cancellation of the imaginary part in the case of higher
statistics in principle, 
we can take the real part of the observables. 
We here use the jackknife method in order to evaluate the statistical errors \cite{JK}.
Since the observables have an imaginary part,  we evaluate the error with complex phase 
and  take the correlation between complex observables and the complex
phase into account. 
Then, we take the real part of the obtained error bars. 

When we evaluate the fourth derivative of $\log Z$ with respect to $\hatmu$, $\CUM4 $, (the function form is like $\expv{\mathcal{O}} = \expv{a} + \expv{b}\expv{c} + \expv{d}^2$),
we generate jackknife samples of each expectation value ($\expv{a}_{\mathrm{bin}}, \expv{b}_{\mathrm{bin}}, \expv{c}_{\mathrm{bin}}, \expv{d}_{\mathrm{bin}} $).
Then, we calculate the jackknife samples of the observable ($\expv{\mathcal{O}}_{\mathrm{bin}} = \expv{a}_{\mathrm{bin}} + \expv{b}_{\mathrm{bin}}\expv{c}_{\mathrm{bin}} + \expv{d}_{\mathrm{bin}}^2$).
Finally we calculate expectation value and error bars of the observable 
by using $\expv{\mathcal{O}}_{\mathrm{bin}}$.
When we calculate the cumulant ratios ($\kappa \sigma^2 = \CUM4 /  \CUM2, S \sigma = \CUM3 /  \CUM2$), 
we use the same technique as well.


\end{document}